\documentclass[pra,aps,twocolumn,groupaddress,longbibliography]{revtex4-1}
\usepackage{latexsym}
\usepackage{lineno}
\usepackage{hyperref}
\usepackage{url}
\usepackage{graphicx}
\usepackage{verbatim}
\usepackage{multirow}
\usepackage{amsmath}
\newcommand{\beq}{\begin{eqnarray}}
\newcommand{\eeq}{\end{eqnarray}}
\usepackage{mathrsfs}
\usepackage{float}
\usepackage[usenames, dvipsnames]{color}
\usepackage{mathtools}
\usepackage{slashed}
\usepackage{graphicx}   
\usepackage{epstopdf}
\usepackage{subfigure}  
\usepackage{hyperref}   
\usepackage{bbold}
\usepackage{wasysym}
\usepackage{amssymb}
\usepackage{feynmp}
\usepackage[symbol]{footmisc}

\usepackage[latin1]{inputenc}
\usepackage{tikz}
\usetikzlibrary{decorations.pathmorphing}
\usetikzlibrary{shapes.misc}
\tikzset{cross/.style={cross out, draw=black, minimum size=8*(#1-\pgflinewidth), inner sep=0pt, outer sep=0pt},
cross/.default={1pt}}
\usetikzlibrary{patterns,math}
\begin{document}

\title{Extended analytical BCS theory of superconductivity in thin films}

\author{\textbf{Riccardo Travaglino}$^{1}$}%
\author{\textbf{Alessio Zaccone}$^{2}$}%
\email{alessio.zaccone@unimi.it}
 
 \vspace{1cm}
 
\affiliation{$^1$ Department of Physics and Astronomy ``A. Righi", University of Bologna, via Irnerio 46, 40126 Bologna, Italy. \\
$^{2}$Department of Physics ``A. Pontremoli'', University of Milan, via Celoria 16,
20133 Milan, Italy.}

\begin{abstract}
We present an analytically solvable theory of BCS-type superconductivity in good metals which are confined along one of the three spatial directions, such as thin films. Closed-form expressions for the dependence of the superconducting critical temperature $T_c$ as a function of the confinement size $L$ are obtained, in quantitative agreement with experimental data with no adjustable parameters. Upon increasing the confinement, a crossover from a spherical Fermi surface, which contains two growing hollow spheres corresponding to states forbidden by confinement, to a strongly deformed Fermi surface, is predicted. 
This crossover represents a new topological transition, driven by confinement, between two Fermi surfaces belonging to two different homotopy classes.
This topological transition provides a mechanistic explanation of the commonly observed non-monotonic dependence of $T_c$ upon film thickness with a maximum which, according to our theory, coincides with the topological transition. 
\end{abstract}
\maketitle

\section{Introduction} 
Superconductivity in spatially confined systems is an important topic both for our basic understanding of superconductivity and quantum matter in general\cite{Ginzburg1968,Bianconi,ScienceAdv,Ovadyahu1973}, as well as for its many technological implications.
Many studies have been devoted, since at least the mid 20th century, to rationalizing the dependence of the superconducting critical temperature $T_c$ on the thin film thickness $L$. While in the early days the superconducting thin films were mostly amorphous\cite{Buckel1954,Buckel1956},
with the advent of modern preparation techniques, also thin films with good crystalline order can be obtained\cite{goodcomp}.

Over several decades, much research has thus been devoted to understanding superconductivity in thin films with amorphous structure\cite{Ovadyahu1973,Ovadyahu2008,Kapitulnik}.
The enhancement of the superconducting critical temperature $T_c$ in thin films has attracted much attention also in terms of theoretical models, but it is clear that this enhancement has a lot to do with the amorphous structure of the films\cite{Ovadyahu1973}, which can strongly affect the phonon physics of the system. Recently, strong coupling theory which accounts for the effects of structural disorder on the phonon density of states has shown that electron-phonon coupling, and hence the $T_c$, can be strongly enhanced by the so-called ``boson peak'' phenomenon induced by disorder in the phonon density of states\cite{Setty2020}.

A different line of research has focused on the sheer effect of confinement (i.e. film thickness $L$) on the superconducting $T_c$ without considering the structural disorder. Early numerical studies\cite{ThompsonBlatt,THOMPSON19636} suggested a possible enhancement of $T_c$ upon decreasing $L$ although a mechanistic explanation has remained elusive. 
More recently, experiments on ordered thin films\cite{goodcomp}, besides the above mentioned regime of enhancement upon reducing $L$, have also highlighted a second regime at very low $L$, where instead the $T_c$ grows with increasing $L$. Recent numerical work\cite{valentinis} has confirmed this picture, and demonstrated indeed the presence of a maximum in the curves of $T_c$ vs $L$.

In this paper, we follow in this second line of research and we develop the first fully analytical theory of confinement effects on superconductivity of thin films. We consider good metals within BCS theory \cite{BCS}, and we ignore effects of structural disorder. The theory is based on analytically describing the effect of confinement on the Fermi surface and on the electron density of states. Analytical closed-form expressions are derived for $T_c$ as a function of $L$ in good parameter-free agreement with experimental data of Ref.\cite{goodcomp}. A new \emph{topological transition} in the available momentum space is predicted to occur at a critical $L_c$ value of thickness, which corresponds to the maximum in $T_c$ vs $L$.
Mechanistic explanations for the trends of $T_c$ vs $L$ are obtained in terms of redistribution of electronic states in momentum space from the interior to the Fermi surface and viceversa, driven by confinement.

\section{Theoretical framework} 
\subsection{Confinement model}
Confinement of a quantum system leads to a change in its fundamental properties, because of the redistribution of accessible states in momentum space. Many models can be formulated, by considering a variety of different boundary conditions (BCs), such as periodic or Dirichlet BCs (the so called "hard walls" BCs). The full wave propagation problem in these models can only be solved numerically, and indeed several studies are available. However, numerical solutions often overshadow the physical mechanisms, so that it is desirable to have analytically tractable theories.

\begin{figure}[ht]
    \centering
    \includegraphics[width=.45\textwidth]{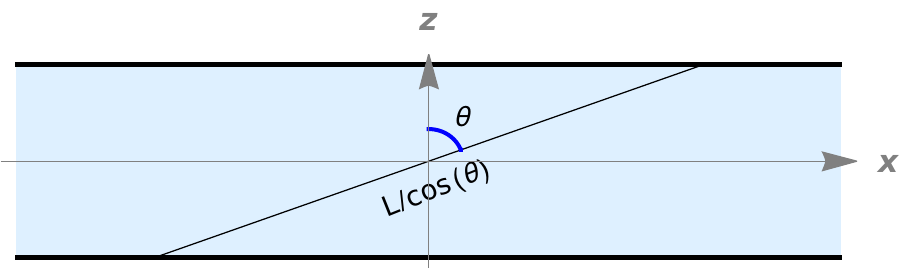}
    \caption{2D section of a thin film of thickness $L$, confined along $z$ and infinite along the $y$ and $x$ directions. A free electron (quantum plane wave) is assumed to have a maximum wavelength equal to the length of the medium in the direction of motion, which can be expressed as a function of the angle $\theta$, thanks to the cylindrical symmetry, as $\lambda_{max} = L/\cos{\theta}$. This leads to a cutoff in the accessible values of wavevector $k$. }
    \label{fig:my_label}
\end{figure}

In this work, following the ideas from Refs. \cite{Travaglino_2022,conf1, conf,Yu2022}, we consider a system confined in the $z$-direction, as shown in Fig. \ref{fig:my_label}, and unconfined in the $x$ and $y$ directions; the following discussion is hence directly relevant to the study of the physics of thin films. In order to perform calculations, electrons are treated as free particles (quantum plane waves) with energy $\epsilon=\frac{\hbar^2k^2}{2m}$, as customary for good metals. The cylindrical symmetry of the system allows one to characterize the states in momentum space of a particle using only the angle $\theta$.

The effect of confinement is taken into account by  imposing a cut-off in the accessible low-energy states, by recognizing that the free quantum particles (or, equivalently, quantum waves) moving in a direction defined by the angle $\theta$ can have a maximum possible wavelength given by~\cite{conf}: $\lambda_{max} = \frac{L}{\cos\theta}$.
This condition implies that the wavelength of a quantum particle cannot exceed the extension of the sample along a particular direction, as clear from Fig.~\ref{fig:my_label}. 
Since the wavelength of a particle is related to its wavenumber by the relation $\lambda=\frac{2\pi}{k}$, this condition is equivalent to a cutoff condition on the minimum possible wavenumber that the free particle can carry: $k_{min} = \frac{2\pi \cos{\theta}}{L}$.
Upon considering plane wave states that propagate in the real-space material depicted in Fig.\ref{fig:my_label}, it is possible to analytically calculate the geometry of the corresponding volume in momentum space. This was done in Refs.\cite{conf,Travaglino_2022} for phonons/elastic waves and bosons, respectively, and the result for phonons is summarized in Fig. 2 of \cite{Travaglino_2022}.
For the example of phonons, as discussed in Ref.\cite{conf}, inside the Debye sphere (i.e. the outer sphere) there are two ``hollow'' spheres of forbidden states in momentum space, both of radius $\frac{\pi}{L}$ centered in $(0,0,\pm \frac{\pi}{L})$. The outer (Debye) sphere, of radius $k_{D}$, represents all allowed states for plane waves in a bulk unconfined material while the two hollow spheres inside represent states that are not accessible due to the confinement. Therefore, when converting sums over wave vectors to integrals over the available momentum space, the integrals must not be carried on the whole Debye sphere, as would be standard for phonons in unconfined materials, but rather on the manifold given by the Debye sphere minus the two spheres of forbidden states.

We should note that, of course, a gas of free electrons or free bosons (e.g. phonons) in a rectangular slab of an internally isotropic material is described by basic quantum mechanics in terms of plane waves, with a wavefunction $\psi \sim \sin(k_{x} x)\sin(k_{y} y) \sin(k_{z} z)$ where e.g. $z$ could be the confined dimension. This form of wavefunction obviously arises from ``hard-wall'' boundary conditions (BCs), i.e. by imposing that the wavefunction vanishes at the boundaries of the box exactly. Additionally, the plane waves in a 3D (internally isotropic) system of {\it any} shape in real space, must satisfy, in $k$-space, the relation:
\begin{equation}
\frac{1}{k^{2}}(k_{x}^{2} + k_{y}^{2}+k_{z}^{2})=1,
\end{equation}
with $|\mathbf{k}|=k=2\pi/\lambda$ the modulus of the wavevector $\mathbf{k}$, and $\lambda$ is the wavelength: see, e.g., p. 493 in Ref.~\cite{Hill} or p. 138 in Ref.~\cite{kittel}. 
While $k_{x}$, $k_{y}$, $k_{z}$ are quantized by the hard-wall BCs, if the number of atoms/quantum particles $N$ is large enough, $k$ can be treated as a continuous variable, see again e.g. \cite{kittel}.
This is fully consistent with $k$ being a continuous variable for large enough number of particles as customary in the context of Fermi gas or Debye model for free particles in a box.

Furthermore, and importantly, the above confinement model has been quantitatively checked and verified in great detail for phonons in ice (both crystalline and amorphous) under nanometric confinement, by means of detailed atomistic simulations and experiments in Ref. \cite{Yu2022} (see in particular the section ``Failure of the hard-wall boundary conditions'' in the Supplementary Information of  \cite{Yu2022} ).

    \label{fig:spheres}

\subsection{The case of free electrons}
In the following we are interested in describing the effect of confinement on free electrons within the typical assumptions of BCS theory \cite{BCS}. Hence we will neglect the above effect of confinement on phonons since the phonons which mediate the Cooper pairing within BCS theory are high-energy optical phonons near the Debye frequency $\omega_{D}$, which is a quantity that depends exclusively on the interatomic bonding and atomic mass and is not affected by confinement. 

The calculation of the available momentum space for large enough $L$ is just the same as for phonons and coincides with the calculation reported in Ref.\cite{conf}, as long as the two spheres are entirely contained inside the outer (Fermi) sphere.
What changes in the case of free electrons is that the maximum energy of allowed states, i.e. the Fermi energy, is not fixed once and for all. This is of course very different from the case of phonons, where $\omega_{D}$ is strictly an insurmountable limit. 
Hence, in the case of phonons, the two spheres of forbidden states can grow, upon decreasing $L$, only up to the point where they are just touching the Debye wavevector $k_D$. They cannot grow any further than that.

In the case of electrons, instead, the Fermi energy in momentum space, $\epsilon_{F}$, is not an insurmountable limit: in this case, upon increasing the confinement, i.e. decreasing $L$, the two spheres of forbidden states can grow beyond the Fermi level. Upon decreasing $L$, the Fermi energy can still grow in order to keep the total number of available states, $N$, constant and conserved. However, the Fermi energy grows, upon decreasing $L$, slower than the size of the two spheres of forbidden states; this is a fact that will be shown quantitatively in Section IV. Therefore, upon further increasing the confinement, at some point, the two spheres must grow beyond the Fermi level and the situation depicted in Fig. \ref{fig:Fig3}(a) is reached. 
This is the only, but substantial, difference with respect to the calculation done in Section III of Ref.\cite{conf} for phonons, which led to predictions for the confinement-dependence of shear modulus of liquids and solids in agreement with experimental data.

In the next section, these considerations will be explored in terms of quantitatively describing how the electron density of states evolves with the confinement.

\section{Density of states}
Since the number of low energy states will be different in the presence of confinement, the density of states (DOS) will have a different structure than the traditional DOS for free electrons. The DOS $g(\epsilon)$ as a function of energy can be expressed as:
\begin{equation} 
 g(\epsilon) = \frac{d}{d\epsilon} N(\epsilon'<\epsilon),
\end{equation}
where $N(\epsilon'<\epsilon)$ is the number of states having energy smaller than $\epsilon$ and $V$ is the volume.



\begin{figure}[ht]
    \centering
\includegraphics[width=.5\textwidth]{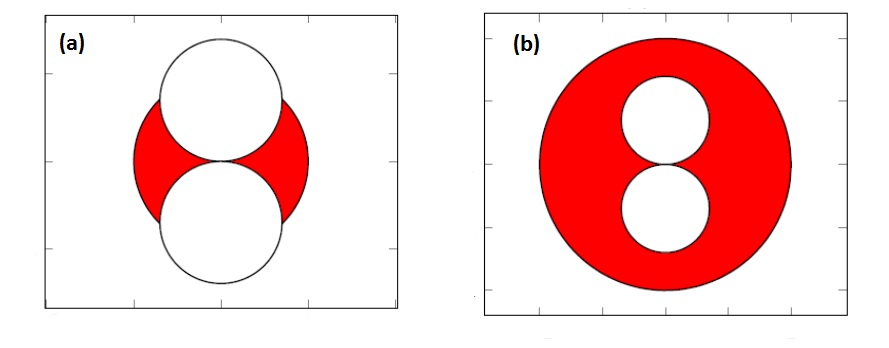}
    \caption{2D section (with reference to e.g. Fig. in the $k_{x}$-$k_{z}$ plane, of the volume of allowed states in $k$-space (red area). The density of states must be calculated considering only the highlighted red zone of available states. In (a), $k<\frac{2\pi}{L}$, and the derivative of the number of states used to evaluate the DOS must take the spheres of forbidden states into account. In (b), $k>\frac{2\pi}{L}$, and therefore the derivative used to evaluate the DOS is unaffected by the forbidden states.}
    \label{fig:Fig3}
\end{figure}

There are two different possibilities depending on whether $k<\frac{2\pi}{L}$ or $k>\frac{2\pi}{L}$ that is, whether the considered sphere in $k$-space crosses or not the two forbidden spheres, see Fig. \ref{fig:Fig3}.

The latter case, depicted in Fig. \ref{fig:Fig3}(b), is the same as it was calculated analytically in Ref.\cite{conf}, and one has
\begin{equation}
    \mathrm{Vol}_{k} = \frac{4}{3} \pi k^3 - 2 \frac{4}{3} \pi \left(\frac{\pi}{L}\right)^3.
\end{equation}
The volume of two forbidden spheres of radius $\frac{\pi}{L}$ is then subtracted from the volume of the Fermi sphere in order to obtain the volume of available states in $k$-space. Since this correction does not depend on $k$, it does not affect the derivative and hence does not affect the DOS. The $g(\epsilon)$ is therefore the usual DOS for free particles:
\begin{equation}
g(\epsilon)= \frac{V (2m)^{3/2}}{2 \pi^2\hbar^3} \epsilon^{1/2}.
\end{equation}

In the former case, namely $k<\frac{2\pi}{L}$, instead, the volume to be considered is shown in Fig.~\ref{fig:Fig3}(a).
This volume can be obtained by subtracting the intersection volume of the red sphere with the two white spheres from the red sphere. 

\begin{figure}
    \centering
    \includegraphics[width=.35\textwidth]{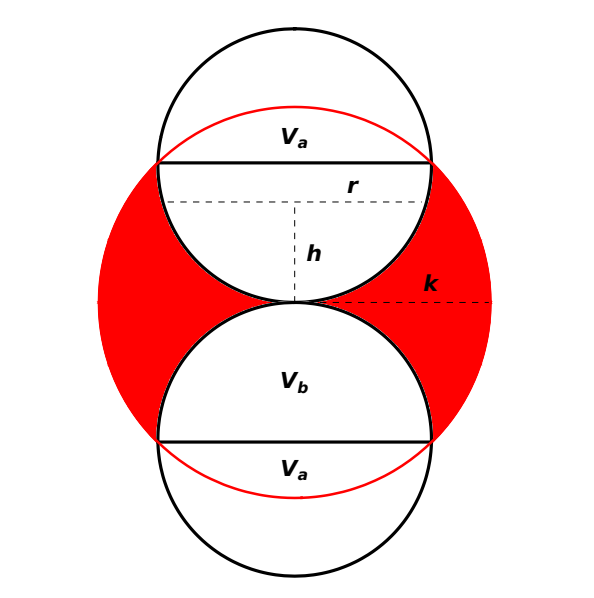}
    \caption{In order to find the density of states for the case $k<\frac{2\pi}{L}$, one has to evaluate the difference between the volume of the red sphere and the volume of the intersection between the red sphere and the prohibited-states white spheres. This latter volume is given by $V_{inter} = 2(V_a + V_b)$, and the two volumes $V_a \mbox{ and } V_b$ can be found by integration of densely infinite stacked circles on horizontal planes. }
    \label{fig:disegnino}
\end{figure}

With reference to Fig. \ref{fig:disegnino}, the total volume of the intersection between the two white spheres of prohibited states and the red sphere can be expressed as $V_{inter} = 2(V_a + V_b)$. The volumes $V_a$ and $V_b$ can be found by integrating, along the $z$ direction, the areas of stacked circles/spherical sections (in general, the volume of a spherical object can be obtained by summing the areas of all the circles that are stacked on each other to form the spherical object; since these circles are densely infinite, the sum is in fact an integral). These circles are identified by the value of variable $h$, which is the distance of the center of the circle from the origin ($k=0$). For the region $a$ (which is just a spherical cap), the radius of these circles can be expressed as $r^2= k^2-h^2$, and the values of $h$ start from $\frac{Lk^2}{2\pi}$, which is, as depicted in Fig. \ref{fig:disegnino}, the plane separating region a from $b$. 

It is then clear that:
\begin{equation}
    V_a = \int_{\frac{Lk^2}{2\pi}}^{k} \pi(k^2-h^2)dh  = \frac{2}{3}\pi k^3 -\frac{k^4L}{2}+\frac{\pi}{3}\frac{L^3K^6}{(2\pi)^3}
\end{equation}
For region $b$, on the other hand, the radii of the circles as function of $h$ are expressed by $r^2 = \left(\left(\frac{\pi}{L}\right)^2-\left(\frac{\pi}{L}-h\right)^2 \right) = -h^2 + \frac{\pi h}{L}$ (because $\frac{\pi}{L}$ is the radius of the confinement-induced sphere).
Therefore the volume of this region can be expressed as:
\begin{equation}
    V_b = \int_0^{\frac{Lk^2}{2\pi}}\pi(-h^2+2\frac{\pi h}{L})dh =  - \frac{\pi}{3}\frac{L^3k^6}{(2\pi)^3} + \frac{Lk^4}{4}
\end{equation}
Hence the total intersection volume is 
\begin{equation}
    V_{inter} = 2(V_a + V_b) = \frac{4\pi k^3}{3} - \frac{Lk^4}{2}
\end{equation}

A simple calculation then yields:
\begin{equation}
    \mathrm{Vol}_k = \frac{4\pi k^3}{3} - V_{inter} = \frac{Lk^4}{2} 
    \label{case2}
\end{equation}
for the total volume of accessible states in $k$-space when $k<\frac{2\pi}{L}$.

Following the same steps as in Ref. \cite{Travaglino_2022}, one finds, for the corresponding DOS:
\begin{equation}
\begin{split}
    N(k'< k) &= \frac{V}{(2\pi)^3}  \frac{L k^4}{ 2 }, \\
    N(\epsilon'< \epsilon) &= \frac{V}{(2\pi)^3}  \frac{L (2m\epsilon)^2}{2 \hbar^4}, \\
    g(\epsilon) &= \frac{d}{d\epsilon} N(\epsilon'<\epsilon) = \frac{V L m^2}{2\pi^3 \hbar ^4 } \epsilon.
    \label{eq:den2}
\end{split}
\end{equation}

Considering the two regimes depicted in Fig. \ref{fig:Fig3}, the overall DOS can be finally expressed as:
\begin{equation}
    g(\epsilon)=\begin{cases}   \frac{V L m^2}{2\pi^3 \hbar ^4 } \epsilon, & \mbox{if }  \epsilon < \frac{2\pi^2 \hbar^2}{mL^2} \\ \frac{ V(2m)^{3/2}}{2\pi^2 \hbar^{3}} \epsilon^{1/2}, & \mbox{if } \epsilon > \frac{2\pi^2 \hbar^2}{mL^2}.
\end{cases}
\label{eq:g}
\end{equation}
In reality it is possible that there is a smooth crossover between the two regimes, which may also depend on the detailed system-specific boundary conditions of the sample and which cannot be determined within the current analytical approach.\\

\begin{figure}[ht]
    \centering
    \includegraphics[width=0.45\textwidth]{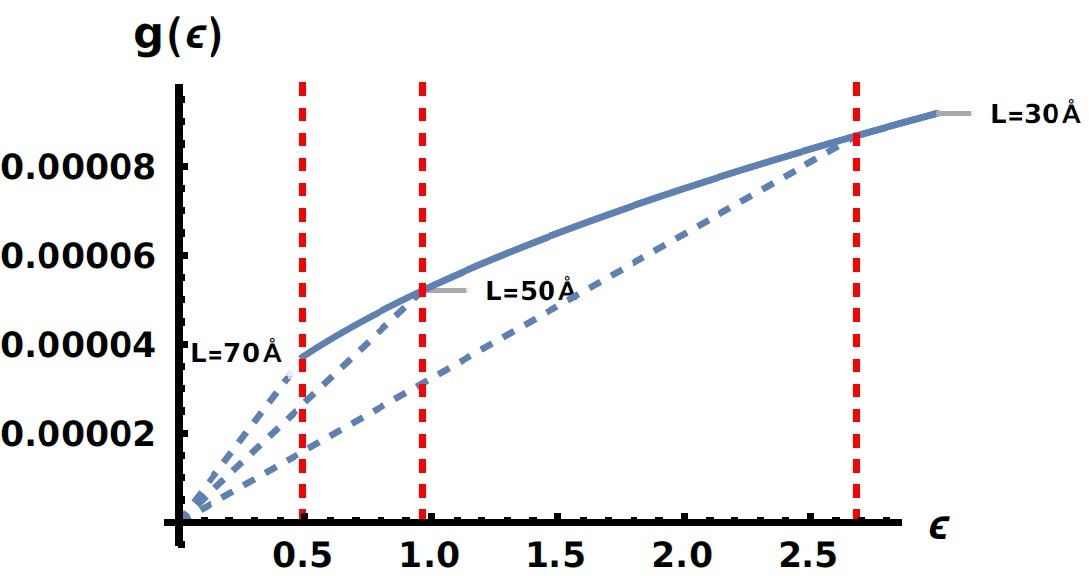}
    \caption{Density of states (DOS) for different values of $L$. The blue, dashed lines represent the linear in $\epsilon$, low energy regime of the DOS, while the blue solid line is the regular free-electron DOS. The crossover between the two forms of the DOS, marked by the vertical red dashed lines, occurs at different values of energy depending on the value of $L$.  As $L$ gets smaller, the region in which the DOS is linear in $\epsilon$ gets wider. (Energy is expressed in
    units of $10^{-20}$ J, and the DOS has consequently units of $10^{20}$ J$^{-1}$).}
    \label{fig:Fig4}
\end{figure}

The form of $g(\epsilon)$ is shown in Fig. \ref{fig:Fig4}. It is clear that, as $L$ increases, the DOS approaches its bulk value for all energy values except for a window that goes from  $\epsilon=0$ up to a crossover energy $\epsilon^{*}$. Furthermore, in order to properly determine the DOS, the spin degeneracy $g_s$ still has to be taken into account as a multiplicative factor. 

In the following, the critical energy $ \frac{2\pi^2 \hbar^2}{mL^2}$, at which the crossover between the two forms of the DOS takes place, will be indicated as $\epsilon^*$. It is interesting to notice that, although the DOS Eq.\eqref{eq:g} is a continuous function, it is not differentiable in the point $\epsilon=\epsilon^*$.

\section{Fermi energy under confinement}
The primary consequence of the crossover between the two regimes in the density of states presented above, is a shift in the Fermi level as a consequence of the fact that some of the low-energy states are prohibited. This becomes particularly important in the following when we will apply the above model to superconductors. Supposing that $\epsilon_F > \epsilon^*$, the Fermi energy can be evaluated by imposing the total number of electrons to be $N$. At $T=0$,
\begin{equation}
    N = \int_0 ^{\epsilon_F} g_{s} g(\epsilon) d\epsilon,
\end{equation}
where  $g_{s}$ accounts for spin degeneracy, and it is $g_s =2$ for electrons.
The integral has to be evaluated in the two regions with the two different DOS regimes:
\begin{equation}
\begin{split}
    N&=  \int_0 ^{\epsilon^*}  2 \ \frac{VL m^2}{2\pi^3 \hbar ^4 } \epsilon  d\epsilon + \int_{\epsilon^*} ^{\epsilon_F} 2 \frac{V (2m)^{3/2}}{(2\pi)^2(\hbar)^3} \epsilon^{1/2} d\epsilon \\
    &=\frac{VL(2m)^2}{(2\pi)^3\hbar^4} \left(\frac{2\pi^2\hbar^2}{mL^2}\right)^2+\\  &+\frac{4V(2m)^{3/2}}{3(2\pi)^2\hbar^3} \left[\epsilon_{F}^{3/2}-\left(\frac{2\pi^2\hbar^2}{mL^2}\right)^{3/2}\right] \\
    &= \frac{4}{3} \frac{V(2m)^{3/2}}{(2\pi)^2 \hbar^3} \epsilon_{F}^{3/2} - \frac{4}{3}\frac{\pi V}{L^3}.
\end{split}
\end{equation}
Here $\epsilon_F$ is the Fermi energy, which can be immediately evaluated explicitly as:
\begin{equation}
\begin{split}
    \epsilon_F &= \frac{\hbar^2}{2m} \left(\frac{N}{V} 3\pi^2 +\frac{2\pi^3}{L^3}\right)^{2/3} \\ 
    &=\frac{\hbar^2}{2m} \left(\frac{3\pi^2 N }{V}\right)^{2/3}\left(1+\frac{2}{3} \frac{\pi}{\rho L^3}\right)^{2/3} \\
    &=\epsilon_F ^{bulk} \left(1+\frac{2}{3} \frac{\pi}{\rho L^3}\right)^{2/3}.
    \label{eq:corrfermi}
\end{split}
\end{equation}

The bulk value is the value obtained using the entire Fermi sphere, and $\rho=N/V$ is the number density of the electrons. This result supports the intuition that the Fermi level is shifted upwards because of the significant number of lower-energy forbidden states due to confinement.  

If the value of $L$ is small enough, it might as well happen that $\epsilon_F < \epsilon^*$: this is exactly the situation anticipated in Sec. II.C, and depicted in Fig. \ref{fig:Fig3}(a). The threshold value $L_c$ where this happens can be found by equating the two energies $\epsilon_F = \epsilon^*$, giving:
\begin{equation}
     \frac{\hbar^2}{2m} \left(\frac{N}{V} 3\pi^2 +\frac{2\pi^3}{L^3}\right)^{2/3} = \frac{2\pi^2 \hbar^2}{mL^2}
     \end{equation}
from which we obtain the confinement size $L_c$ at which $\epsilon_F < \epsilon^*$, as
\begin{equation}
L_c \equiv \left(\frac{2\pi}{\rho}\right)^{1/3}.
\end{equation}

When $L$ is smaller than this value, the Fermi energy can be found through the new condition for $N$:
\begin{equation}
    N =  \int_0 ^{\epsilon_F}  2 \frac{VL (2m)^2}{(2\pi)^3 \hbar ^4 } \epsilon d\epsilon = \frac{VL(2m)^2}{(2\pi)^3\hbar^4}\epsilon_F^2 
\end{equation}
from which we obtain
\begin{equation}
\epsilon_F = \frac{\hbar^2}{m}\left[\frac{(2\pi)^3\rho}{ L}\right]^{1/2}.
  \label{eq:Ef2}
\end{equation}
Therefore the total Fermi energy can be expressed as a piecewise function:
\begin{equation}
    \epsilon_F = \begin{cases} \epsilon_F ^{bulk} \left(1+\frac{2}{3} \frac{\pi}{\rho L^3}\right)^{2/3} \mbox{ if } L > L_c = \left(\frac{2\pi}{\rho}\right)^{1/3} \\ \\
     \frac{\hbar^2}{m}\left[\frac{(2\pi)^3\rho}{ L}\right]^{1/2} \mbox{ if } L < L_c = \left(\frac{2\pi}{\rho}\right)^{1/3}
    \end{cases}
\end{equation}
for the two different regimes.

\section{BCS gap equation in confined superconductors}
Equipped with the quantum confinement model described above, we can now proceed to the analytical implementation of the constraints of confinement within the BCS theory of superconductivity, with the goal of determining the $T_c$ as  a function of the confinement size $L$. 

In our notation, $U_{\Vec{k}\Vec{k'}}$ is the attractive phonon-mediated potential for Cooper pairing, which, in the BCS framework, takes the form:
\begin{equation}
    U_{\Vec{k}\Vec{k'}}= \begin{cases}  -U, & \mbox{if }  |\epsilon-\epsilon_F| < \epsilon_D \\ 0, & \mbox{otherwise }
    \end{cases}
\end{equation}
where $\epsilon_D \equiv \hbar \omega_D$ is the Debye energy, with $\omega_D$ the Debye frequency of the solid. The phonon modes just below $\omega_{D}$ are high-frequency optical modes related to vibrations of few atoms and it is assumed that their frequencies, and the corresponding phonon DOS near the Debye level, are not altered by the confinement.

Using the Bogoliubov method to account for finite-temperature effects, one obtains\cite{Rickayzen,Parks}
\begin{equation}
     \Delta_{\vec{k}} = -\sum_{\vec{l}} U_{\vec{k}\vec{l}} \frac{\Delta_{\vec{l}}}{2E_{\vec{l}}}\tanh\left(\frac{\beta E_{\vec{l}}}{2}\right)
\end{equation}
where the sum is performed in the shell around the Fermi energy of width given by the Debye energy. This leads to \cite{BCS}:
\begin{equation}
    \frac{1}{g(\epsilon_F)U} = \int_0^{\beta_c\epsilon_D/2} \frac{\tanh(x)}{x}dx = \mbox{ln}(1.13\beta_c\epsilon_D),
\end{equation}
where $\beta_c$ indicates the critical value for $\beta=1/k_{B}T$, namely the value at which the superconducting transition occurs.
Inverting the relation yields \cite{BCS}:
\begin{equation}
    k_B T_c = 1.13 \epsilon_D \exp\left[-\frac{1}{g(\epsilon_F)U}\right].
\label{eq:t_c_bcs}
\end{equation}
Equation \eqref{eq:t_c_bcs} provides the closed-form BCS expression to estimate the critical temperature at which the material becomes a superconductor. 

It is clear that the result for the energy gap in the ground state $\Delta(T=0)$ depends on the form of the density of states and on the Fermi level, both of which will be changed due to confinement effects. These changes are considered in the next subsections.

\subsection{The regime $L > L_c$ and $\epsilon^* < \epsilon_F$}
  In the regime for which $\epsilon^* < \epsilon_F$, and hence $L > L_c$, the relevant DOS is the usual $g(\epsilon) = g_s \frac{(2m)^{3/2}}{(2\pi)^2\hbar^3} \epsilon^{1/2} $, which has to be evaluated at the Fermi level corrected for confinement, given by Eq.\eqref{eq:corrfermi}. It is worth noting that, while the confinement changes the value of the Fermi energy, it does not affect the Debye energy which is a property dictated solely by the atomic structure and bonding of the material.
  In this regime, for the density of states at Fermi level we obtain:
  \begin{equation}
      g(\epsilon_F) = 2\frac{ V(2m)^{3/2}}{2\pi^2\hbar^3} \epsilon_F^{1/2} = g^{bulk}(\epsilon_{F}) \left(1 + \frac{2}{3} \frac{\pi}{\rho L^3}\right)^{1/3} 
 \label{Fermi_1}
  \end{equation}
  
The latter expression can be substituted in the BCS gap equation,  $\Delta(T=0) = 2\epsilon_D \exp{[-\frac{1}{U g(\epsilon_F)}]}$, to obtain a gap equation corrected for confinement:
  \begin{equation}
      \Delta(T=0) =  2\epsilon_D \exp{\left(-\frac{1}{U g^{bulk}(\epsilon_{F}) \left(1 + \frac{2}{3} \frac{\pi}{\rho L^3}\right)^{1/3} }\right)}.
  \end{equation}

  The critical superconducting temperature $T_c$ is related to the gap $\Delta$ via\cite{BCS} $T_c \simeq \frac{2 \Delta}{3.52 k_B}$, leading to the following expression that accounts for confinement:
  \begin{equation}
      T_c = \frac{4\epsilon_D}{3.52 k_B}  \exp{\left(-\frac{1}{U g^{bulk}(\epsilon_{F}) (1 + \frac{2}{3} \frac{\pi}{\rho L^3})^{1/3} }\right)} \label{eq:t_c}
  \end{equation}
  This is a key result of this paper.
  Equation \eqref{eq:t_c} expresses the critical temperature corrected for the effects of confinement, which  thus departs from the bulk (unconfined) value.
  This relation shows that, according to this confinement model, the $T_c$ should increase as the value of $L$ is decreased. Confinement effects become relevant when the inverse cube of the thickness $L$ of the superconducting thin film becomes comparable to the number density of electrons, which has values of around $n\approx 10^{28}$ m$^{-3}$; therefore confinement should start to have an appreciable effect on critical temperature for $L\sim 10^{-9}$ m.   
  
Therefore, the critical temperature of superconducting thin films should be greater than for bulk superconductors. The increase remains rather small until one gets to sufficiently small values of $L$ where the increase in $T_c$ with respect to he bulk becomes significant. If the value of $L$ gets too small, however, some of the approximations used above are not valid anymore and one has to resort to different expressions valid for the regime $L < L_c$, that will be developed in the next section.


\subsection{The regime $L < L_c$ and $\epsilon^* > \epsilon_F$}
The discussion in the previous section is valid only if $L>L_c=\left(\frac{2\pi}{\rho}\right)^{1/3}$. In the opposite case of $L< L_c$, the two forbidden-states spheres cross the Fermi level, which implies that the Fermi surface is no longer spherical, and new important effects have to be taken into account. We will still assume that the DOS can be approximated by using its value at the Fermi energy: this is justified by the fact that the corrected DOS is continuous at the critical point, thus it can be Taylor expanded around $\epsilon_F$ with sufficient precision, assuming that $\epsilon_D \ll \epsilon_F$ as it is the case in metals.

The Fermi energy takes the form shown in Eq.\eqref{eq:Ef2}, and the relevant DOS is given by equation Eq.\eqref{eq:den2}.  Proceeding as above, we find
\begin{equation}
\begin{split}
    g(\epsilon_F) &= 2\frac{V L (2m)^2}{(2\pi)^3 \hbar^4}  \frac{\hbar^2}{m}\left((2\pi)^3 \frac{\rho}{L}\right)^{1/2} \\
    &= 2\frac{V m \sqrt{L \rho}}{\sqrt{2} \pi^{3/2}\hbar^2}.
\end{split}\label{Fermi_2}
\end{equation}
The DOS at Fermi level thus appears to have a completely different dependence on $L$ than in the previous case, which will eventually result in a different dependence of $T_c$ on $L$, as we shall see.
This then leads to a different formula for the critical temperature $T_c$, valid in the regime $L<L_c =\left(\frac{2\pi}{\rho}\right)^{1/3}$:
  \begin{equation}
      T_c = \frac{4\epsilon_D}{3.52 K_B}  \exp{\left(-\frac{1}{U g^{bulk}(\epsilon_F)}\frac{(3\pi^2 \rho)^{1/3}}{\sqrt{2\pi L \rho}}\right)}.
      \label{eq:Tc2}
  \end{equation}
 This is another key result of this paper. As a first sanity check, we note that the factor $\frac{(3\pi^2 \rho)^{1/3}}{\sqrt{2\pi L \rho}}$ inside the exponential is correctly dimensionless.
 In the present case, the weak-coupling approximation $\sinh{x}\approx e^x/2$ is even more justified, since for $L\rightarrow 0$ the argument of the hyperbolic sine tends to infinity. 
 It should be noticed that it has been assumed that the general BCS equation for the $T_c$ remains the same, while just the Fermi energy and the electronic DOS are different in the two regimes. This approximation is the same as requiring that the DOS can be approximated as a constant value in the whole range of interaction, and thus can be taken out from the integral\cite{BCS}. In principle, the fact that the spheres of forbidden states intersect the shell on which integration is carried out, should change all the procedure. A way to improve the calculations, without making use of such approximation, will be considered in the Appendix \ref{section:exact_calc}. With this more precise calculation we have shown that the above approximation amounts to neglecting an additional term in the expression for $T_c$, which however turns out to be very small, i.e. contributing a correction just about 0.7\% of the actual $T_c$.
This demonstrates the robustness of the approximation.
 
 \section{Results and discussion}
 \subsection{Theoretical predictions}
\begin{figure}[ht]
    \centering
\label{fig:Tca}{\includegraphics[width=.45\textwidth]{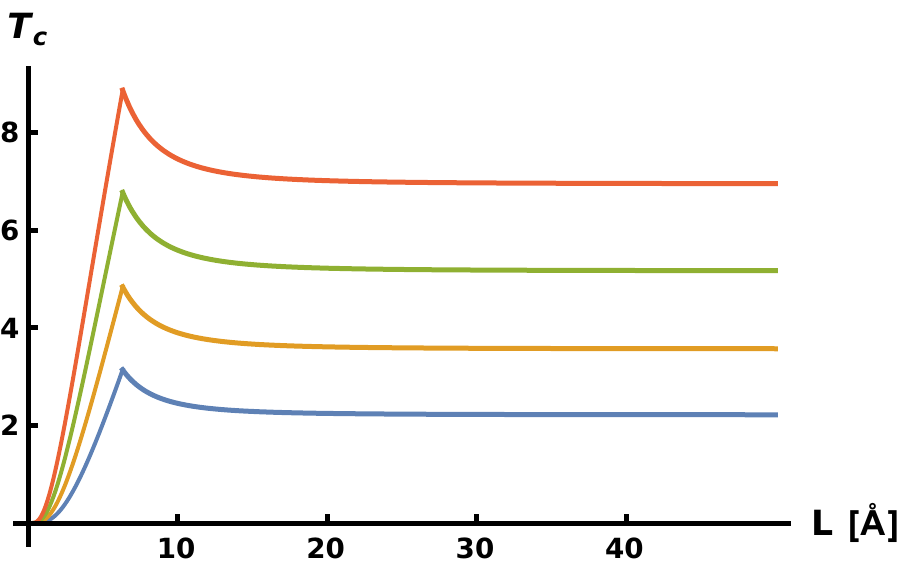}} \quad
\label{fig:Tcb}{\includegraphics[width=.45\textwidth]{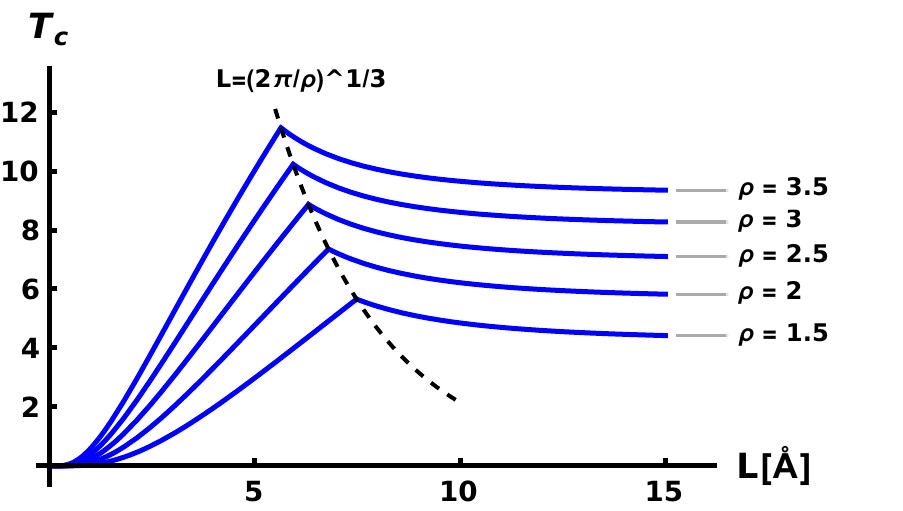}}
 \caption{Critical temperature $T_c$, in Kelvin, as a function of the  film thickness $L$, in \AA ngstrom. Top figure: critical temperature $T_c$ vs $L$ plotted for different values of $U$. The energy density at fixed density of states at Fermi level is fixed $g(\epsilon_F) = 1 eV^{-1}$, with values of $U$ ranging from 0.3 to 0.6 $eV$.
 Bottom figure: critical temperature $T_c$ vs $L$ plotted as the density of electrons $\rho$ is varied while keeping $U$ fixed. $\rho$ is expressed in units of $10^{28} m^{-3}$.}
 \label{fig:critical_temperature_bcs}
  \end{figure}

Calculations based on the above theoretical model for $T_c$ are shown in  Fig. \ref{fig:critical_temperature_bcs}, upon varying the Cooper pairing strength $U$ (top panel) and upon varying the electron density $\rho$ (bottom panel).
 In the limit of large $L$ values, the critical temperature $T_c$ is expected to remain essentially constant until the inverse cube of the film thickness $L$ reaches a few nanometers. At the crossover between the two regimes discussed in the previous sections, namely $L_c=\left(\frac{2\pi}{\rho}\right)^{1/3}$, there is a  maximum in the $T_c$ vs $L$ curves. Upon approaching the maximum from above, the $T_c$ increases of about 10\% with respect to the bulk value. Upon further decreasing $L$ below the peak, a fast descent to zero sets in, in the regime where the distorted Fermi surface of Fig. \ref{fig:Fig3}(a) has to be taken into account. The fact that the critical temperature drops to zero implies that superconductivity should be destroyed by taking vanishingly small films. It should be noted that the cusp at the peak (discontinuity of the first derivative) is due to the approximations made in the previous sections. If the whole DOS (instead of the constant-DOS approximation) were considered in the solution of the gap equation, then a slight rounding effect  would remove the cusp at the peak, as shown in the Appendix \ref{section:exact_calc}. 
 
Finally, we also note that the electron-phonon coupling constant, in the simplest BCS estimate, is given by $\lambda = U g(\epsilon_{F})$. Since there is no reason to expect the attractive interaction energy $U$ to depend on $L$, it is then clear that, according to our model, the dependence of $\lambda$ on $L$ is entirely dictated by the dependence of $g(\epsilon_{F})$ on $L$. In particular, $\lambda$ will first increase with increasing film thickness $L$, according to Eq. \eqref{Fermi_2}, it will display a maximum at $L_{c}$ and then it will decay to its bulk value with further increase in $L$ according to Eq. \eqref{Fermi_1}.
 
\subsection{Comparison with experimental data and mechanistic interpretation}
Experiments on thin superconducting films have been conducted for more than fifty years now, given the many technological applications and the fundamental importance for a deeper understanding of the phenomena. In earlier days, the study of the dependence of critical temperature on film thickness $L$ was made more difficult by the fact that the critical temperature depends strongly on the disorder in the material, which is higher in thin films since it was hard to preserve crystalline order up to small thickness. The experiments conducted in the literature show clearly that the properties of the superconducting film depend on a number of factors, particularly on the thickness, on the degree of thermal equilibrium of the film, the type of substrate, the degree of development of the surface, and the texture\cite{russi}. 

Improvements in technological capabilities in more recent years have allowed for production of atomically ordered thin films down to only two atomic layers in the $z$ direction. Recent state-of-art experiments conducted in Ref.\cite{goodcomp} on lead films supported on silicon substrates, show clearly that the critical temperature remains almost constant, with very small oscillations (much smaller than those predicted by the old Thompson-Blatt model~\cite{ThompsonBlatt}) in correspondence of the atomic layers, until the film becomes only a few atomic layers thick. Then there is an increase of $T_c$ upon further decreasing $L$ followed by a sudden drop as the thickness reaches two atomic layers (interestingly, no data is present for films of exactly three atomic layers, since this condition is thermodynamically unstable). 

\begin{figure}[ht]
    \centering
    \includegraphics[width=0.45\textwidth]{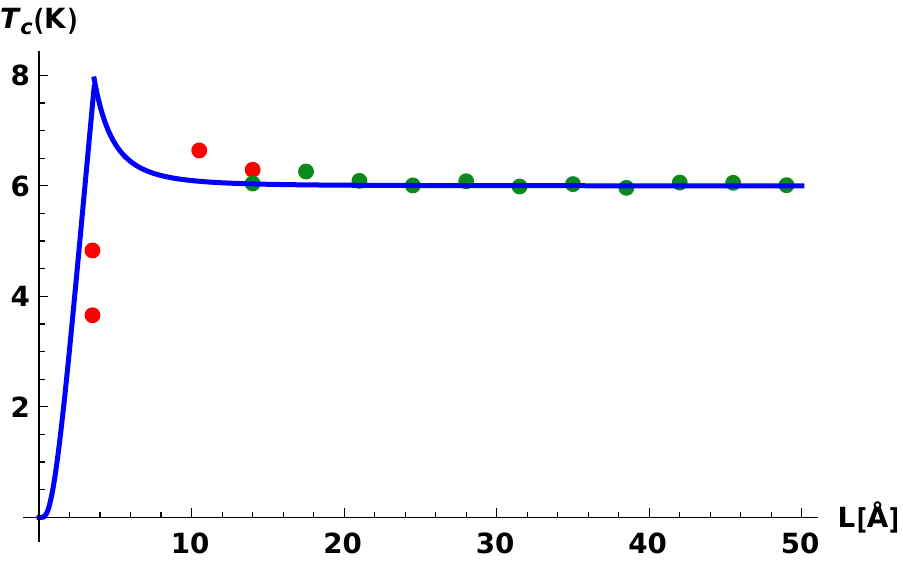}
    \caption{Semi-quantitative comparison between the theory prediction of Eqs. \eqref{eq:t_c} and \eqref{eq:Tc2} (solid lines) with experimental data (circles) of Pb thin films. Red circles are experimental results from Refs. \cite{goodcomp,goodcomp2}. Green and red points correspond to two different data sets, and the two measurements with the lower $L$ correspond to two values of $T_c$ measured for two films made of two layers but crystallized in two different ways. The theoretical fitting was done using appropriate bulk values for Pb, in particular
  $U=2.68 \cdot 10^{-19} J$ (or $1.67~eV$), which was found by inverting Eq.\eqref{eq:t_c_bcs} with $T_c= 6K$ as the ``bulk'' value (since this is the experimental large-$L$ limit), while the electron density was taken as $\rho = 13.2 \cdot 10^{28} m^{-3}$, the DOS at Fermi level in the bulk $0.125 \cdot 10^{19}~J$ or $0.2~(eV)^{-1}$, and the Debye energy as $\epsilon_D = 105 K \cdot k_B$ as tabulated in Ref. \cite{kittel}.}
    \label{fig:goodcomp}
\end{figure}

The comparison of the theoretical model with the experimental data of Ref.\cite{goodcomp} is shown in Fig. \ref{fig:goodcomp}. The experimental systems are thin films of lead (Pb), synthesized via epitaxial growth in crystalline form, according to Ref. \cite{goodcomp}, and with thickness ranging from 15 to 2 atomic layers.
We recall that Pb is a type-I conventional superconductor, qualitatively well described by BCS theory. In the bulk, the superconducting critical temperature is $T_{c}=7.2~K$, the energy gap $\Delta = 1.365 ~meV$, Fermi energy $\epsilon_{F}^{bulk}=9.37~eV$, DOS at Fermi level $g(\epsilon_{F}^{bulk})=0.2~(eV)^{-1}$ and Debye energy $\epsilon_D=105K \cdot k_{B}$ as tabulated\cite{kittel}.

The comparison shows a good agreement between analytical theory (Eqs. \eqref{eq:t_c} and \eqref{eq:Tc2}) and experiments, with practically no adjustable parameter:  all the material-specific parameters entering Eqs. \eqref{eq:t_c} and \eqref{eq:Tc2}, i.e. $\rho$, $\epsilon_D$ and $\epsilon_{F}^{bulk}$ have been taken equal to their tabulated values for Pb, with the exception of the pairing energy $U$, which turns out to be equal to $1.67~eV$ from the fitting, i.e an acceptable value (corresponding to an electron-phonon coupling constant $\lambda\approx 0.33$ in the bulk), safely lower than the Fermi energy, $\epsilon_{F}^{bulk}=9.37~eV$. The bulk critical temperature was taken at the asymptotic value towards which the experimental points tend in the large $L$ limit, and from this value the value of $U$ was found by inverting the BCS equation, Eq. \eqref{eq:t_c_bcs}.

In particular, the theory is able to capture both the $T_c$-enhancement induced upon increasing confinement, at larger $L$, i.e. in the regime $L>L_c = \left(\frac{2\pi}{\rho}\right)^{1/3}$, as well as the trend at lower $L$, in the regime $L<L_c = \left(\frac{2\pi}{\rho}\right)^{1/3}$, where $T_c$ decreases with further decreasing $L$.\\

Importantly, this crossover is explained by our theory in terms of the crossover from the regime ($L>L_c$) where the Fermi surface is still perfectly spherical and the two spheres of forbidden states induced by the confinement are comprised within the Fermi sphere and grow with decreasing $L$, Fig. \ref{fig:Fig3}(b), to the regime ($L<L_c$) where the Fermi surface gets strongly distorted by the excluded volume of forbidden states. \\

In the first regime, $L>L_c$, the DOS at Fermi level $g(\epsilon_F)$ increases with decreasing $L$ because the excluded volumes of the two spheres of forbidden states become bigger, while the Fermi radius $k_F$ remains constant. Hence, since $N=const$, the states that are excluded from the two forbidden spheres need to be accommodated in the remaining allowed volume (red volume in Fig. \ref{fig:Fig3}(b)), which includes the surface of the outer sphere i.e. the Fermi surface. Therefore, the number of states that are on the Fermi surface has to increase, which means that $g(\epsilon_F)$ has to increase with decreasing $L$.

At $L_c$, a crossover takes place into the regime of Fig. \ref{fig:spheres}(a) where, due to confinement, the two spheres of forbidden states are now intersecting the Fermi surface, which therefore is no longer spherical and gets strongly distorted. Upon further decreasing the thickness $L$, the DOS at Fermi level $g(\epsilon_F)$ now decreases due to the increasing deformation of the Fermi surface caused by the confinement, with the appearance of a necking point at the center of the original Fermi sphere. In this regime, as the two spheres of excluded states grow in size upon reducing $L$, the specific surface of the allowed volume (red in Fig. \ref{fig:spheres}(b)) grows, which means that less states are accommodated on the surface as $L$ decreases. 

\subsection{Topological transition at $L = L_c$}
It is evident from Fig. \ref{fig:Fig3} that as soon as the two spheres of  forbidden  states  intersect  the  original  Fermi  sphere (red), a new topological transition takes place:  as $L$ crosses the critical value $L_c$, the space of accessible states transitions to a new, topologically distinct space: while for $L > L_c$ the space is simply connected, for $L \leq L_c$ the homotopy group changes, since it becomes (isomorphic to) $\mathbb{Z}$, considered as a group with the standard operations. This suggests that the drastic change in the trend of the $T_c$ as a function of confinement at the critical point $L_c$ is a consequence of this change in the topological structure of the accessible states in $k$-space.

Even though this topological transition is somewhat reminiscent of a Lifshitz transition as it is found in metals at high pressures\cite{Lifshitz,Volovik}, its nature is quite different. In this case, there is no apparent necking and there is no apparent way by which the Fermi surface can be split into two disconnected domains.

The prediction of a topological transition induced by confinement is a key prediction of this work, which is certainly worth further investigating in the future.
For example, recent work \cite{Chamon} has provided a systematic classification of topological transitions in Fermi surfaces of different materials under different conditions, which are associated with singularities in the density of states. This was done, for strictly 2D systems, by means of catastrophe theory. The same approach can be used in future work to better understand and classify the new topological transition discovered here for thin films in the context of 3D and quasi-2D materials.

\subsection{Comparison with previous numerical works}

Many different mathematical models have been formulated in order to describe the effect of confinement in superconducting thin films, all of which require numerical calculations. The models generally differ from each other depending on the boundary conditions which are taken into account in the numerical calculations, which clearly determine completely the solutions of the gap equation.

A classical example is that of the Thompson-Blatt model, which considers the case of ``hard walls'' boundary conditions (BCs), that is by imposing the wavefunctions to be exactly zero at the surface of the film\cite{ThompsonBlatt}. This leads to a discretization of energy levels, and as a consequence to an oscillation of the critical temperature as a function of the width of the film, as shown in their original paper\cite{ThompsonBlatt}. 

Similar strong oscillations were also reported in later numerical works such as Ref.\cite{yu-strongin}.

These models predict unphysically large oscillations of $T_c$ which are (for low values of $L$) as large as 70-80\% of the bulk value. Moreover, the effect of confinement is supposed to become relevant for values of $L$ much larger than the few angstroms that are shown in Fig. \ref{fig:goodcomp}. Recently, numerical progress on smoothening of these oscillations has been obtained in Ref.\cite{valentinis}.

It has to be noted, however, that the trend shown by experiments on a large class of superconductors is that oscillations of $T_c$ as a function of film thickness are either below the level of experimental noise or much smaller than the oscillations reported in numerical studies that used hard-wall BCs. Hence, the model proposed in this work could provide a better fit of experimental results by not having to implement hard-walls BCs.

\section{Conclusion}
We presented a theory of superconductivity which extends BCS theory to superconducting thin films by fully and analytically taking into account how the electronic density of states, and the Fermi surface, get modified and distorted as a function of confinement in one spatial direction.

The key effect driving the enhancement of $T_c$ upon increasing the confinement (i.e. upon decreasing the film thickness $L$) in a regime of sufficiently large $L$ is provided by the redistribution of electron states in the allowed $k$-space. The confinement contributes two hollow spheres (corresponding to excluded or forbidden states due to the cutoff on wavelengths caused by confinement) inside the Fermi sphere. As confinement grows, the two spheres grow causing a redistribution of states in the remaining $k$-space volume, which  includes the Fermi surface. As a consequence, the density of states (DOS) on the Fermi surface increases with increasing confinement, and the $T_c$ increases.
As the two spheres of forbidden states keep growing, they eventually intersect the Fermi surface, which causes a Lifshitz-type necking phenomenon right at the center of the original Fermi sphere. Hence the Fermi surface is no longer spherical and gets now strongly distorted by the confinement.
Now the specific (Fermi) surface of the volume of available states increases as the excluded-states spheres grow, due to the distortion. As a consequence of the increased specific surface in momentum space, the available states get redistributed from the Fermi surface into the inner volume in $k$-space (red in Fig. \ref{fig:Fig3}(a)), hence the DOS at Fermi level now decreases with increasing the confinement.

Therefore the typical non-monotonic dependence of $T_c$ on film thickness $L$, with a maximum, is mechanistically explained by the theory in terms of the confinement-induced distortions of the available momentum space and redistribution of states between the latter and the Fermi surface.

The theory produces closed-form analytical expressions for the $T_c$ as a function of the various physical parameters (as in BCS theory) including the confinement size $L$.
These expressions provide good agreement with experimental data with no adjustable parameters.

The concepts and mechanistic understanding developed based on the theory presented here can be used in future work to guide the design of materials with improved superconducting properties and enhanced $T_c$.

Furthermore, a new topological transition in momentum space is predicted at a critical value of thickness $L_c$, with the topology of the available momentum space belonging to two different homotopy classes at $L < L_c$ and at $L > L_c$, respectively. This topological transition is entirely induced by increasing the confinement and presents different features compared to the well known Lifshitz transition in metals.
This new topological transition driven by confinement is a very interesting topic in its own right, to be further explored in future studies.

Another important direction for future work will be to extend the present theory to the strong-coupling regime, possibly in connection with numerical implementation of the full Migdal-Eliashberg theory\cite{Margine1,Margine2}, which will enable application of the theory to amorphous films\cite{Ovadyahu2008}.\\

\textit{Acknowledgements} We thank Dr. Chandan Setty for reading an early version of the paper and for useful comments. This work has received funding from the European Union (ERC, ``Multimech'', contract no. 101043968) and from the US Army Research Office through contract nr. W911NF-22-2-0256. 

\section*{Data availability}
The data that support the findings of this study are available from the corresponding authors upon reasonable request.

\appendix
\section{Calculations of Sec. V.B by accounting for the full DOS }
 \label{section:exact_calc}
\begin{figure}[ht]
      \centering
      \includegraphics[width=0.45\textwidth]{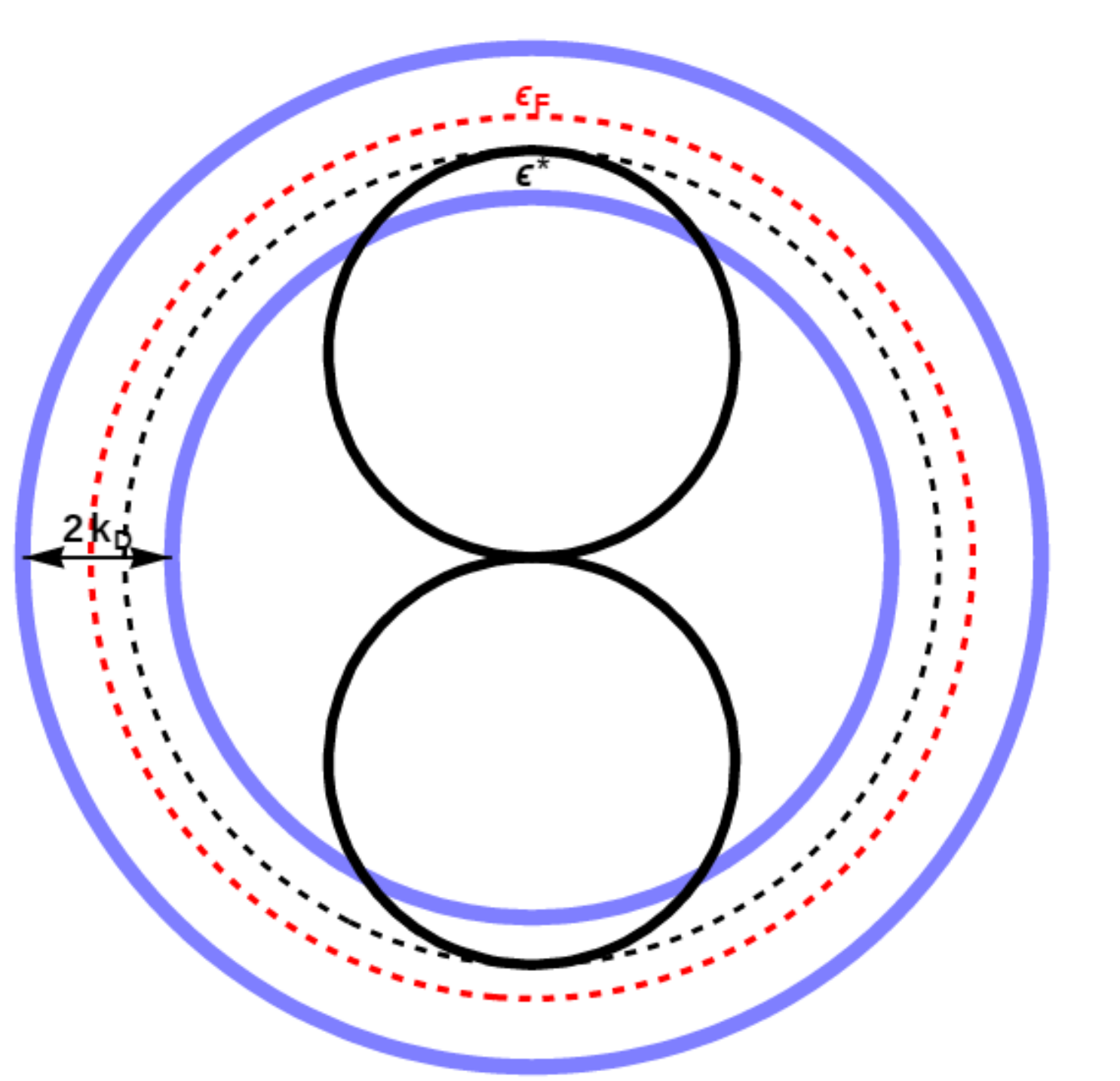}
      \caption{If the cutoff spheres of forbidden states intersect the sphere identified by $\epsilon_F - \epsilon_D$, then the approximations of Sec. V.B need to be further verified.}
      \label{fig:ex_calc}
  \end{figure}

 We present how the calculations of Section V.B get modified if one considers the full DOS.
 If the situation is that shown in Fig. \ref{fig:ex_calc}, namely if we have $\epsilon^* \in (\epsilon_F - \epsilon_D, \epsilon_F + \epsilon_D)$, exact calculations involve considering the full density of states (DOS), hence solving the equation: 
 \begin{widetext}
\begin{equation}
    1 = \frac{U}{2} \int_{-\epsilon_D}^{\epsilon_D} g(E) \frac{1}{\sqrt{\Delta^2+E^2}}  dE   =\frac{U}{2}\left[\int_{-\epsilon_D} ^{\epsilon^*-\epsilon_F} \frac{VLm^2}{2\pi^3\hbar^4} \frac{E+\epsilon_F}{\sqrt{\Delta^2+E^2}}dE 
    +\int_{\epsilon^*-\epsilon_F} ^{\epsilon_D} \frac{V(2m)^{\frac{3}{2}}}{(2\pi)^2\hbar^3}\frac{\sqrt{E+\epsilon_F}}{\sqrt{E^2+\Delta^2}}dE\right]
    \label{eq:exact}
\end{equation}
 \end{widetext}
  
  The limit of integration is $\epsilon^*- \epsilon_F$ because the integration variable is the energy measured with respect to the Fermi level. Although the first term in Eq. \eqref{eq:exact} can be exactly solved, this is not true for the second term, which cannot be integrated in closed form. Therefore, the only way to consider the second integral is by making some approximations, for example expanding to the first order the square root in the numerator of the integral. Solving the two integrals separately gives:
  \begin{widetext}
  \begin{equation}
     \frac{U}{2} \int_{-\epsilon_D} ^{\epsilon^*-\epsilon_F} \frac{VLm^2}{2\pi^3\hbar^4} \frac{E+\epsilon_F}{\sqrt{\Delta^2+E^2}}dE  
      = \frac{UVLm^2}{4\pi^3\hbar^4}\left[\sqrt{\Delta^2+(\epsilon^*-\epsilon_F)^2} +
      \epsilon_F \mbox{arcsinh}\left(\frac{\epsilon^*-\epsilon_F}{\Delta}\right)
      -\sqrt{\Delta^2+\epsilon_D^2} + \epsilon_F \mbox{arcsinh}\left(\frac{\epsilon_D}{\Delta}\right)
      \right]
  \end{equation}
  \end{widetext}
  The second integral is solved by considering the approximation $\sqrt{E+\epsilon_F} = \sqrt{\epsilon_F\left(1+\frac{E}{\epsilon_F}\right)} \approx \sqrt{\epsilon_F} (1+\frac{E}{2\epsilon_F})$, which is justified by the fact that this calculation only considers $E \in (-\epsilon_D, \epsilon_D)$, hence $E \ll \epsilon_F$. This approximation makes calculations much simpler, since the simplified integrals are the same (except for prefactors) as those solved above:\\
\begin{widetext}
\begin{equation}
\begin{split}
&\int_{\epsilon^*-\epsilon_F}^{\epsilon_D} \frac{UV(2m)^{\frac{3}{2}}}{2(2\pi)^2\hbar^3}\frac{\sqrt{E+\epsilon_F}}{\sqrt{E^2+\Delta^2}}dE  \approx \int_{\epsilon^*-\epsilon_F}^{\epsilon_D} \frac{UV(2m)^{\frac{3}{2}}}{2(2\pi)^2\hbar^3}\frac{\epsilon_F}{\sqrt{E^2+\Delta^2}}dE +\\
&+\int_{\epsilon^*-\epsilon_F}^{\epsilon_D} \frac{UV(2m)^{\frac{3}{2}}}{2(2\pi)^2\hbar^3} \frac{1}{s\sqrt{\epsilon_F}}\frac{E}{\sqrt{E^2+\Delta^2}}dE
= \frac{UV(2m)^{3/2}}{8\pi^2\hbar^3}\sqrt{\epsilon_F} \left[ \mathrm{arcsinh}\left(\frac{\epsilon_D}{\Delta}\right) - \mathrm{arcsinh}\left(\frac{\epsilon^*-\epsilon_F}{\Delta}\right) \right] +\\
 &+ \frac{UV(2m)^{3/2}}{16\pi^2\hbar^3\sqrt{\epsilon_F}} \left[\sqrt{\epsilon_{D}^{2}+\Delta^2} - \sqrt{\Delta^2+(\epsilon^*-\epsilon_F)^2} \right].
 \end{split}
 \end{equation}
 \end{widetext}
 
 It is clear that summing the two integrals, and inverting the total equation found in this way to find the value of the gap, cannot be done analytically. 
 It is instructive to first consider the case of $\epsilon^* = \epsilon_F = \frac{2\pi^2\hbar^2}{mL^2}$, hence the case for which the two spheres of forbidden states are tangent to the Fermi sphere. This happens for $L=L_c=\left(\frac{2\pi}{\rho}\right)^{1/3}$.
 In this case, the two integrals above (referred to as $I_1$ and $I_2$, respectively) are evaluated as follows: 
 \begin{gather}
     I_1 = \frac{UVLm^2}{4\pi^3\hbar^4} \left[ \Delta - \sqrt{\Delta^2 + \epsilon_D}\right] + \frac{UVm}{2\pi\hbar^2L}\mathrm{arcsinh}\left(\frac{\epsilon_D}{\Delta}\right) \\ 
    I_2 = \frac{UVm}{2\pi\hbar^2L}\mathrm{arcsinh}\left(\frac{\epsilon_D}{\Delta}\right) + \frac{UV m^2 L}{8\pi^3\hbar^4}\left[\sqrt{\epsilon_D^2+\Delta^2}-\Delta\right].
 \end{gather}
 The total gap equation can be obtained by summing the two expressions, and equating to 1, as in Eq. \eqref{eq:exact}. It is thus easy to obtain:
 \begin{equation}
     1 = \frac{UVm}{\pi \hbar^2 L} \mathrm{arcsinh}\left(\frac{\epsilon_D}{\Delta}\right) + \frac{UVLm^2}{8\pi^3\hbar^4}\left[\Delta-\sqrt{\Delta^2+\epsilon_D^2}\right]
 \end{equation}
 
 The second term is a correction to the calculations performed in the previous sections, while the first term leads to the same results seen before, evaluating the gap at $L_{c}=\sqrt[3]{\frac{2\pi}{\rho}}$ (so that we have $\epsilon_F = \epsilon^*$).
 
 Graphic solutions of this equation show that this correction term is very small, i.e. just about 0.7\% of the critical temperature value. In fact, the error that is made by not considering this term is smaller (with the parameters used) than the error which comes from the approximation $\sinh{x} \approx \frac{e^x}{2}$, which is made in the assumption of weak coupling\cite{cohen}. Since the value of $\epsilon^*$ lies in a small range around the Fermi energy, the correction will also be small (for continuity reasons) for values for which $\epsilon_F = \epsilon^* +\delta\epsilon$.
 Therefore, this shows that performing exact calculations without supposing the density of states to be constant in the range of interaction provides a correction to the results obtained in the Section V.B which is negligible, and does not alter the overall structure of the solutions shown in Fig. \ref{fig:critical_temperature_bcs}. In fact, this approximation was already used and discussed in \cite{BCS}. Hence, the approximations made in the previous sections are \emph{a posteriori} justified.\\
 
 \begin{figure}[ht]
     \centering
     \includegraphics[width=0.45\textwidth]{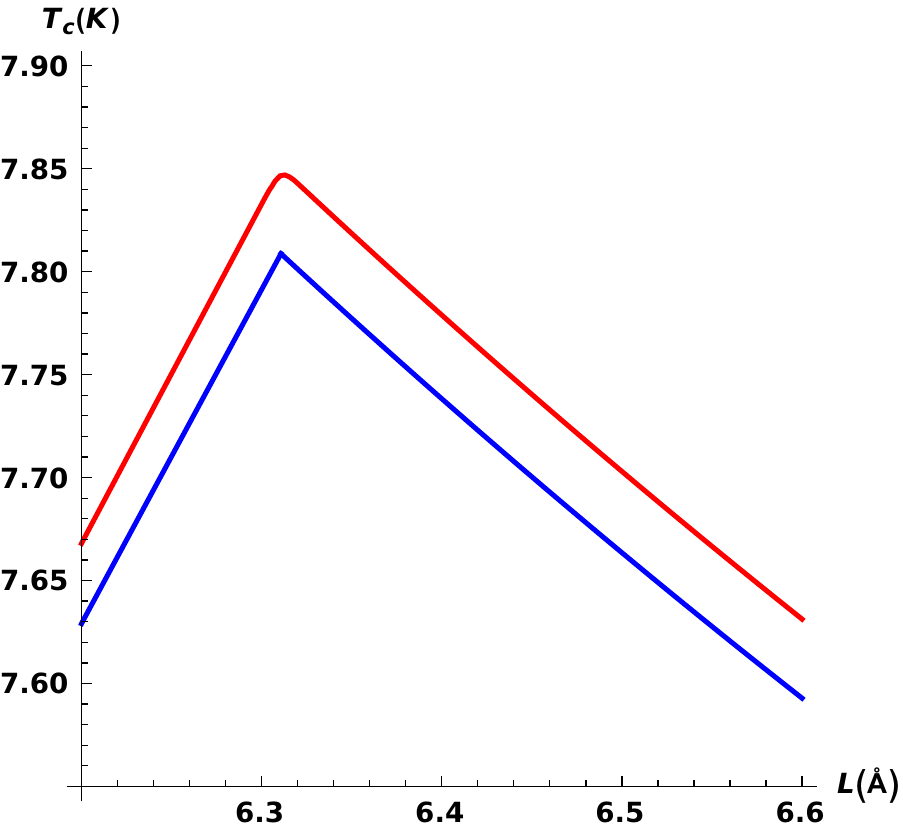}
     \caption{Zoomed-in comparison between the analytical approximation calculation of Sec. V.B and the numerical more accurate computation of this Appendix. The red curve is the curve obtained by numerical solution of the integrals in Eq.\eqref{eq:exact}, and the blue curve is the approximate solution of Sec. V.B. The main reason in the slight rise in the critical temperature is actually due to the fact that the approximation $\sinh{x} \approx \frac{e^x}{2}$ was not made; the effect of considering the whole density of states is responsible for the rounding of the curve around the critical value, $L = (\frac{2\pi}{\rho})^{1/3}.$ From the plot it is clear that the approximations made are completely justified also in confined superconductors. }
     \label{fig:exact_comp}
 \end{figure}
 
Finally, we show in Fig. \ref{fig:exact_comp} a direct comparison between the $T_c$ vs $L$ curve computed using the constant DOS approximation of Sec. V.B, and the less approximate computation outlined above in this Appendix and obtained via numerically solving Eq.\eqref{eq:exact}.
 Using the full DOS leads to rounding off the cusp at $L=L_c$ and to a very slight overall increase of the predicted $T_c$. The latter effect is due to relaxing the (weak-coupling) approximation $\sinh{x} \approx \frac{e^x}{2}$ used in the analytical treatment of Sec. V.B.

 \bibliography{Bib}

\end{document}